\documentclass{article}

\usepackage{arxiv}

\usepackage[utf8]{inputenc} % allow utf-8 input
\usepackage[T1]{fontenc}    % use 8-bit T1 fonts
\usepackage{hyperref}       % hyperlinks
\usepackage{url}            % simple URL typesetting
\usepackage{booktabs}       % professional-quality tables
\usepackage{amsfonts}       % blackboard math symbols
\usepackage{nicefrac}       % compact symbols for 1/2, etc.
\usepackage{microtype}      % microtypography
\usepackage{lipsum}
\usepackage[pdftex]{graphicx} 
\usepackage{amsmath}
\usepackage{graphicx}
\usepackage[section]{placeins}
\usepackage{multirow}
\usepackage{multicol}
\usepackage{algorithm}
\usepackage{algorithmic}
\usepackage{caption}

\usepackage[square,numbers]{natbib}
\title{Rock, Paper, Scissors, ... Dynamite \\
\hfill \\
\large A Model of Disruption from New Technologies}

\author{
  Andrew J. Lohn\\
  Center for Security and Emerging Technology, Georgetown University\\
  Washington, DC, USA \\
  \texttt{drew.lohn@georgetown.edu} \\
}

\begin{document}
\maketitle

\begin{abstract}
We seek to understand the effect of adding disruptive highly-capable new technologies to competitions by assessing the addition of Dynamite to Rock-Paper-Scissors. We find that providing a versatile Dynamite move to only one player provides limited value (win probability increases from 50\% to 55.5\%) and is played rarely. That value decreases further if the game is expanded beyond just the original three moves. We also observe several mechanisms by which prior moves can become strategically unplayable, or obsolete. We hope that this model illustrates some non-intuitive aspects of developing new versatile technologies. We also hope that it illustrates some pitfalls for developers and integrators to avoid in order to create value rather than merely capability. 
\end{abstract}

% keywords can be removed

\section{Introduction}
\label{sec:introduction}

Children sometimes disrupt the game Rock-Paper-Scissors by adding a new move with new capabilities, such as Dynamite. All agree that Dynamite beats Rock but it is less clear whether Scissors can snip the wick, and whether Paper snuffs the flame or burns. "The Theoretical Throws Bureau of the World RPS Society has made a clear position that Dynamite is incompatible with the RPS trinity relationship and its inclusion will alter the cycle of connection the game is supposed to form." \cite{WRPSAdynamite} In our case though, it is precisely that alteration of the game that we seek to understand. Our interest is not in the game of Rock-Paper-Scissors, nor in Dynamite the noun. Our interest is in understanding Dynamite the adjective, as in "that new AI model or joint strike fighter is dynamite." 

Disruptive new technologies are arriving at an accelerating pace, especially related to, or resulting from, AI. Ahead of their arrival, there are too few models or mechanisms for anticipating their impact and for intuiting about an increasingly tumultuous world. We recently set out to account for the various ways that AI can provide competitive advantage, for example through information asymmetries, first-mover advantages, or by providing more materiel.\cite{lohnAIcompetition} One of AI's potential advantages comes from providing more moves or options to choose from. AI itself could be an additional option in the sense that an employer, software developer, military commander, or other decision maker could select an independent AI system or model where they would have previously selected a human or more traditional tool or process. Alternatively, AI might make existing tools or humans more capable, providing decision makers with more capable or versatile options to choose from. This paper considers the effect of adding more capable new moves in the simplified setting of Rock-Paper-Scissors. \cite{nashPy}

Models are necessarily oversimplifications that can be "intellectually useful, but only at the most elementary level." \cite{schellingElementary} Simple analogies allow people to reason about complex issues. They also provide a baseline to depart from when considering the ways that the world is disanalogous from those simple models. For example, if a model suggests that a new technology provides less value than anticipated, that does not necessarily mean that the technology provides little value. Rather, the model clarifies what needs to be done to avoid low-value outcomes.

In Rock-Paper-Scissors, competitors select moves that are either useful or vulnerable depending on their opponent's choice. There are countless competitions where opponents have arsenals of options that each defeat some opponent moves while being defeated by other countermoves. A steady state can arise when competitors iteratively develop moves and countermoves in pursuit of an advantage or an acceptably balanced stasis. \cite{healeyStasis, trenches, schmidtDrones} Competitors may not be satisfied with a balance and may seek to disrupt it by introducing new technologies. But those new technologies, including the most capable AI systems, still have shortcomings and vulnerabilities to exploit. \cite{lohnHackingAI,nistAIvulns,lohnBrittleness}

We are motivated by the impact of AI in this paper, but the findings are not restricted to AI. Perhaps a new fighter jet is equipped for stealth bombing, close air support, counter-air assault, and suppression of enemy air defenses. In a sports setting, a player may create matchup problems by scoring from multiple positions, or by being able to defend many opponents. In a corporate setting, a company may staff itself to prepare for uncertain circumstances. Whether an employee, team, tool, or approach is a good choice, a bad choice, or a neutral choice for a particular project or task might depend on the actions of market competitors. In all cases, an option that is capable over a wider range of circumstances may be valuable and may disrupt aspects of the competition.

\subsection{Summary of Results}

The simple model of adding Dynamite to Rock-Paper-Scissors can illustrate several competitive dynamics. We consider the original three-move game as well as expanded games with more moves. We also consider Dynamite of varying capability, by which we mean the number of opposing actions it defeats. We explore cases where both competitors have access to the new additional Dynamite move, cases where only one has access to it, and cases where both have access but to Dynamite moves of differing capabilities.

The fundamental competitive dynamics that we illustrate with this game structure are listed below.

\begin{itemize}
    \item A new move may not change the expected outcomes or add any competitive value even if it changes strategies, alters competition dynamics, and makes prior moves unplayable.

    \item When the new move is valuable, that value is lower than might be expected, even for very capable moves.

    \item Increasing capability (as in the number of opposing moves that a new move defeats) has decreasing marginal value. That value saturates such that further increases in capability provide no additional value. 
    \begin{itemize} 
        \item On the other hand, decreasing the number of moves it draws against continually increases its value.
    \end{itemize}

    \item The high-capability new move is played rarely in optimal strategies. Its existence forces the opponent to adjust their strategy in ways that are more vulnerable to less capable moves which are then played more often as a result.
    \begin{itemize}
        \item Usage rates might not reflect value. Companies that use or sell AI might need alternative models to determine and assign value rather than relying on how often it is used.
    \end{itemize}

    \item A new high-capability move can make other moves unplayable in several ways:
    \begin{itemize}
        \item A move may be dominated, either weakly or strictly, such that the new move is never worse or always better. In the Rock-Paper-Scissors structure, the original moves are rarely dominated by new high-capability moves.

        \item The addition of a new move can create new optimal strategies that do not include all prior moves, even ones that are not dominated. This is prevalent in the Rock-Paper-Scissors structure. This is a less-appreciated mechanism for obsolescence and is worth considering in the context of workforce disruptions. 

        \item Many moves can remain playable despite being nearly dominated by the new move. In fact, their usage can increase dramatically, even if the only opponent move they defeat is also defeated by the new additional move. 
    \end{itemize}

    \item We consider cases where competitors each add new highly-capable moves that are not equivalent. Competitors can have enhanced-Dynamite, limited-Dynamite, or be entirely Dynamite-denied. 
    \begin{itemize}
        \item In a competition where one player has enhanced Dynamite, providing their opponent with limited Dynamite can make some moves unplayable for both players that the original higher-capability Dynamite alone does not. The unplayable moves can be different for each player even if Dynamite defeats identical moves.

        \item The limited-Dynamite can be more valuable than the enhanced-Dynamite despite defeating fewer options and losing to the same number. That is because the introduction of a new more capable moves breaks the symmetry of the game such that some moves become more important than others. A limited new move may achieve high value if the few moves it defeats have become more important.
    \end{itemize}
\end{itemize}

\section{Dynamite and Dominance}
\subsection{Impervious Dynamite}
Clearly if Dynamite is impervious in the sense that it defeats all three of Rock, Paper, and Scissors while losing to none, then it should be played in every round. If only one competitor has that additional move then they win every round. If both competitors have it, then every round becomes a Dynamite-Dynamite draw. 

The game is completely upended in one sense and completely unchanged in another. It shifts from a fair balance of wins and losses to a game of certain draws. If your interest in the game is the expected value it can return, then the addition of impervious Dynamite causes no change at all. If your interest is in round-to-round variability, then it is unrecognizable after the addition of an impervious option.

\subsection{Weak Domination}
If Dynamite loses to two, namely both Scissors and Paper, then Dynamite is "weakly dominated" by Paper and should never be played. The game is not affected at all by the addition of a weak option that is dominated by existing options.

If Dynamite defeats two and loses to one then it is not dominated. Specifically, if it explodes Rock and burns Paper but is snipped by Scissors, then Dynamite weakly dominates Paper instead of the other way around. Paper becomes obsolete. The game becomes Rock-Scissors-Dynamite with the same dynamics as Rock-Paper-Scissors. This transition is illustrated by the payoff matrix in Tables \ref{RPSD}. Given that Paper is removed from the game, an equivalent way to view the addition of Dynamite is that it is an expansion of Paper's capabilities rather than a separate additional move that overtakes Paper. As an analogy for technology, perhaps AI enhances some existing tool or user's abilities so that they are more capable over a wider range of circumstances.

\begin{table}[!htb]
    \centering
    \caption{The payoff matrix for Rock (R), Paper (P), Scissors (S), and Dynamite (D) becomes just Rock, Scissors, Dynamite when Paper is weakly dominated by Dynamite.}
    \label{RPSD} 
    
    \begin{tabular}{c|c|c|c|c|}
        & S & P & R & D \\ \hline
        S & 0 & 1 & -1 & 1 \\ \hline 
        P & -1 & 0 & 1 & -1 \\ \hline
        R & 1 & -1 & 0 & -1 \\ \hline
        D & -1 & 1 & 1 & 0 \\ \hline
    \end{tabular}

    \textbf{$\downarrow$}
    
    \begin{tabular}{c|c|c|c|}
        & S & R & D \\
        \hline
        S & 0 & -1 & 1 \\ \hline 
        R & 1 & 0 & -1 \\ \hline
        D & -1 & 1 & 0 \\ \hline
    \end{tabular}
\end{table}

As long as Dynamite does not defeat everything, then the World Rock Paper Scissors Association is wrong that the addition of Dynamite alters the trinity relationship or game dynamics. Whether Dynamite loses to both Paper and Scissors or only to Scissors, the game is unchanged, albeit with the possible obsolescence of one of the original moves. Although their official decision was mostly humorous, it illustrates how easily even an organization that is intimately familiar with a contest that is primitively simple can still misjudge the impact of new additions.

This simple case also illustrates one type of obsolescence, where an option is removed because it never outperforms the new addition, although a weakly-dominated option is not necessarily non-viable the way a strictly-dominated one is. This simple model also shows how even when options are dominated, the competition's dynamics may remain unchanged. It also shows how the competition's dynamics can change without changing the overall outcome; adding impervious Dynamite changes strategies and the certainty of outcomes without changing the expected payoff. 

Subsequent sections discuss how the competitive dynamics do change more substantially if only one competitor has access to Dynamite and we consider larger contests than the three-move Rock-Paper-Scissors.

\section{Dynamite Denial}
If Dynamite defeats both Rock and Paper, and if only one competitor has access to Dynamite, then the dynamics shift. Paper is still dominated for the Dynamite-holder but remains useful for the Dynamite-denied. The payoff matrix is illustrated in Table \ref{DvP}

\begin{table}[!htb]
    \centering
    \caption{Payoff matrix for asymmetric access to Dynamite.}
    \label{DvP}
    \begin{tabular}{c|c|c|c|}
    & S & R & P \\ \hline
    S & 0 & -1 & 1 \\ \hline 
    R & 1 & 0 & -1 \\ \hline
    D & -1 & 1 & 1 \\ \hline
\end{tabular}
\end{table}

This game can be easily solved to give the equilibrium strategies for both competitors as well as the expected value of asymmetric access to Dynamite. The Dynamite-holder wins one extra game in every nine for a 55.6\% win probability. Notably, dynamite does not deliver an outsized percentage of those wins. Rather, the threat of Dynamite forces the Dynamite-denied competitor to use Scissors more often, which are vulnerable to the Dynamite-holder's Rock. The Dynamite-holder wins by using Rock more often while keeping their Dynamite usage unchanged. The optimal strategies are shown in Table \ref{DvPstrats}.

\begin{table}[!htb]
    \centering
    \caption{The competitor who is Dynamite-denied increases Scissor use, which the Dynamite-holder exploits with Rock.}
    \label{DvPstrats}
    
    Dynamite-Denied \\
    \begin{tabular}{|c|c|c|}
    \hline
    $p_S$ & $p_R$ & $p_P$ \\ \hline 
    4/9 & 3/9 & 2/9 \\ \hline
    \end{tabular} 

    %Janky way to make a space between these two tables that I'm trying to keep together.
    \begin{tabular}{c}
    \end{tabular}

    Dynamite-Holder \\
    \begin{tabular}{|c|c|c|}
    \hline
    $p_S$ & $p_R$ & $p_D$ \\ \hline 
    2/9 & 4/9 & 3/9 \\ \hline
    \end{tabular}
\end{table}

That the Dynamite-holder increases their Rock usage rather than Dynamite is familiar in an athletic context where opposing coaches commonly double-team the best player to "make someone else beat us." Coaches might be surprised though that the possession of a Dynamite player only increases their win probability by such a small amount: from 50\% to 55.6\%. That small win percentage decreases further in competitions with more than three moves, as will be discussed in the next section.

\section{Expanded Rock, Paper, Scissors}
\label{sec:Expanded}
Most interesting competitions have more than three options, but there are many ways to expand the Rock-Paper-Scissors. Each option could defeat one, lose to one, and draw against all others. We will call this a one-to-one competition, where one possible arrangement is for each move to defeat the previous move's countermove, as illustrated in Table \ref{RPSexpandedOtoO}. Alternatively, each option may defeat several options, lose to several options, and draw against itself. An example many-to-many contest is the famous Rock, Paper, Scissors, Lizard, Spock expansion where each option defeats two and loses to two. \cite{rpsLS} That payoff matrix is shown in Table \ref{RPSexpandedMtoM}.

\begin{table}[!htb]
    \centering
    \caption{Payoff matrices for expanded one-to-one and many-to-many competitions.}
    \label{RPSexpanded} 

    One-to-One\\
    \label{RPSexpandedOtoO}
    \begin{tabular}{c|c|c|c|c|c|}
        & scissors & paper & $b_{3}$ & $b_{4}$ & rock \\ \hline
        Rock & 1 & -1 & 0 & 0 & 0 \\ \hline 
        $a_{2}$ & 0 & 1 & -1 & 0 & 0 \\ \hline
        $a_{3}$ & 0 & 0 & 1 & -1 & 0 \\ \hline
        $a_{4}$ & 0 & 0 & 0 & 1 & -1 \\ \hline
        Paper & -1 & 0 & 0 & 0 & 1\\ \hline
    \end{tabular}

%Janky way to make a space between these two tables that I'm trying to keep together.
    \begin{tabular}{c}
    \end{tabular}

    Many-to-Many \\
    \label{RPSexpandedMtoM}
    \begin{tabular}{c|c|c|c|c|c|}
        & rock & paper & scissors & spock & lizard \\ \hline
        Rock & 0 & -1 & 1 & -1 & 1 \\ \hline 
        Paper & 1 & 0 & -1 & 1 & -1 \\ \hline
        Scissors & -1 & 1 & 0 & -1 & 1 \\ \hline
        Spock & 1 & -1 & 1 & 0 & -1 \\ \hline
        Lizard & -1 & 1 & -1 & 1 & 0 \\ \hline
    \end{tabular}
\end{table}

In these expanded competitions, the new move's capabilities and weaknesses can vary more widely; it can defeat or lose to few or many opponent moves. We restrict our analysis to new moves that lose to a single opponent move, but we consider new moves that defeat varying numbers of opponent moves. 

The addition of Dynamite also breaks the symmetry of these games. Prior to Dynamite, every move is equivalent to all others. After Dynamite, the utility of moves changes according to whether they defeat Dynamite, draw against it or lose against it. The usefulness of a Dynamite-holder's moves also change according to whether they defeat opposing moves that Dynamite also defeats or whether they defeat opposing moves that Dynamite cannot defeat or loses against.

\subsection{Dominance in Expansion}
In one-to-one contests, as long as Dynamite is defeated by an opponent move, it can only ever dominate one move--the only move that is also defeated by the move that defeats Dynamite. 

In many-to-many contests, there can be many moves that also lose to the opponent move that defeats Dynamite. The number of moves that Dynamite dominates depends on the particular selection of which other opponent moves it defeats. In a many-to-many contest that alternates wins and losses as in Rock-Paper-Scissors-Lizard-Spock, Dynamite that loses to Rock, draws against Paper and defeats all three of Scissors, Lizard, and Spock dominates nothing. Dynamite that defeats everything except the Rock that it loses to dominates two of the five moves, namely Lizard and Scissors.

As we will see in the next section, that does not necessarily imply that the addition of a new Dynamite move makes few moves unplayable or obsolete.

\section{Dynamite Denial in an Expanded One-to-One Competition}
\subsection{Defining Dynamite}
\label{definingDynamite}
In this section, we solve the expanded one-to-one configuration described in Section \ref{sec:Expanded}. We arrange the capabilities of Dynamite as illustrated in Table \ref{kParam}. In that arrangement, there is a Dynamite-holder whose moves prior to Dynamite are $\{a_1, ..., a_n\}$. There is a Dynamite-denied competitor whose moves are $\{b_1, ..., b_n\}$. Dynamite loses to the first opponent move ($b_1$), draws against opponent moves $\{b_2, ..., b_{k-1}\}$ and defeats opponent moves $\{b_k, ..., b_n\}$. Dynamite dominates the Dynamite-holder's last move ($a_n$), so the Dynamite-holder's set of moves becomes $\{a_1, ..., a_{n-1}, a_d\}$.

\begin{table}[!htb]
    \centering
    \caption{The capability of Dynamite is limited by the number of moves it does not beat $k-1$. The table shows Dynamite's payoffs against each opponent option $b_i$.}
    \label{kParam}
    \begin{tabular}{c|c|c|c|c|c|c|c}
           & $b_1$ & $b_2$ & ... & $b_{k-1}$ & $b_k$ & ... & $b_n$ \\ \hline
          $a_d $& -1 & 0 & ... & 0 & 1 & ... & 1 \\ \hline
    \end{tabular}
\end{table}

\subsection{Tactically Useful, Strategically Unplayable}
We derive the Nash equilibrium strategies for this scenario in Appendix \ref{appendixExpandedOtoO} along with the value that the new move provides. The value depends on how many moves are initially available ($n$), how many moves the high-capability move defeats ($n-k+1$), and how many it does not defeat ($k-1)$ as shown in equation \ref{valueOtoO}.

\begin{equation}
\label{valueOtoO}
    v = \frac{2(m-k)}{km(m-k+2)} 
\end{equation}

In equation \ref{valueOtoO}, we use $m$ rather than $n$ because not all of the total available moves are strategically playable. For a given $k$, increasing $m$ initially increases the value $v$, but subsequent increases lead to a peak beyond which increasing the number of moves reduces value. We define $m$ as the optimal number of moves for the Dynamite-holder. They should play all $k-1$ moves that target opponent moves which Dynamite cannot defeat. They should also play $j=m-k+1$ non-Dynamite moves that target opponent moves which Dynamite can defeat, provided that there are enough total moves ($n$) for those to be dismissed. The number of useful moves is given in equation \ref{usefulMoves}.

\begin{subequations}
\label{usefulMoves}
\begin{gather}
    j = \sqrt{2k} \\
    m = k + \sqrt{2k} - 1
\end{gather}
\end{subequations}

Those dismissed moves are tactically useful in the sense that they are not dominated by Dynamite, but they are strategically dominated in the sense that including them reduces the odds of victory. They are technically "outside the support." In the context of obsolescence, there are two perspectives. On the one hand, $\sqrt{2k}$ moves remain strategically useful despite Dynamite defeating the only opposing move that they defeat. On the other hand, if there are many available moves ($n$), a large number of them could be made irrelevant despite appearing useful in that they are not dominated by the addition of the new more capable move. That suggests that AI might make more tools or tasks obsolete than might be anticipated while allowing others to remain useful (or even increase in usefulness) despite appearing redundant.

\subsection{Dynamite's Value in Large Competitions}
Increasing the capability of a new Dynamite move by expanding the number of opponent moves that it defeats (increasing $n$) increases the new move's value as long as $n<m$. Beyond that point, expanding Dynamite's capability does nothing to increase its value because those additional opponent moves are outside the support. Alternatively, increasing the capability of Dynamite by reducing the number of opponent moves that it does not defeat (reducing $k$) does increase its value. 

To build intuition for large contests with many available moves ($n>m$) and many moves that Dynamite does not defeat ($k>>1$), we come to a simple approximation. The value of a new move falls quadratically with the number of opponent moves that it does not defeat as shown in equation \ref{valLargeKmain}.

\begin{equation}
\label{valLargeKmain}
    v \approx 2k^{-2}
\end{equation}

\subsection{Dynamite Usage Rates}
As in the smaller Rock-Paper-Scissors, the value does not come from playing the high-capability move. The optimal distribution of moves is given in equation \ref{probsOtoO}.

\begin{subequations}
\label{probsOtoO}
\begin{gather}
    \label{optimalNonDefeat}
    p_i = \frac{2m+2i(m-k)}{km(m-k+2)}, i\le k-1\\
    \label{optimalDefeat}
    p_i = \frac{2(m-i)}{m(m-k+2)} , k-1 < i < m\\
    \label{optimalDynamite}
    p_d = \frac{2}{k(m-k+2)} 
\end{gather}
\end{subequations}

Without Dynamite, each move would be played with the same probability equal to $1/n$. If the introduction of Dynamite makes moves strategically unplayable ($n>m$) then the probability of playing Dynamite can potentially be higher than any move would be in the absence of Dynamite. But the probability of Dynamite is still less than or equal to the average of the remaining moves. It often has nearly the lowest usage rate. 

Its usage is highest when Dynamite is maximally capable (i.e. $k=2$). In that case, there is only one move that Dynamite does not defeat--the one it loses against. Then the optimal probability for using Dynamite ($p_d$) becomes $1/m$ as it was for the three-move Rock-Paper-Scissors. As $k$ increases, the fraction of Dynamite usage falls. Using the same large-contest approximations ($n>m, k>>1$), equation \ref{usageFalls} shows how the optimal usage of Dynamite ($p_d$) falls compared to the average move ($p_{avg}$).  

\begin{subequations}
\label{usageFalls}
\begin{gather}
    p_d \approx \frac{2}{k^{3/2}} \\
    p_{avg} = \frac{1}{k+k^{1/2} +1} \approx \frac{1}{k}
\end{gather}
\end{subequations}

In the context of AI, these low usage rates could complicate accounting for value or interfere with approaches to pricing. An organization might expect to measure the value of its new technology by how often it is used, but its value might come from how much it alters a competition instead. A double-teamed basketball player may create open shots for others without ever touching the ball. And AI companies may expect to capitalize on its value by charging per use, but if it is rarely used then they will struggle to recover their costs even if it provides significant value. Organizations may need some other methods or simple models to understand, assess, and assign the value of AI creating new moves or adding capability to existing moves in a competitive context.

\section{Expanded Many-to-Many RPS}
Rather than fully solve a many-to-many case, we will present an example to illustrate similar dynamics as were solved for in the expanded one-to-one case. In Rock-Paper-Scissors-Lizard-Spock, each move defeats two, loses to two, and draws against itself. We provide each player with Dynamite that defeats three, loses to one, and draws against itself and one other as shown in Table \ref{DynLizSpock}

\begin{table}[!htb]
    \centering
    \caption{Payoff matrix for Dynamite in Rock-Paper-Scissors-Lizard-Spock}
    \label{DynLizSpock}
    \begin{tabular}{c|c|c|c|c|c|c|}
        & rock & paper & scissors & spock & lizard & \textbf{dynamite} \\ \hline
        Rock & 0 & -1 & 1 & -1 & 1 & -1 \\ \hline 
        Paper & 1 & 0 & -1 & 1 & -1 & 0 \\ \hline
        Scissors & -1 & 1 & 0 & -1 & 1 & -1 \\ \hline
        Spock & 1 & -1 & 1 & 0 & -1 & -1\\ \hline
        Lizard & -1 & 1 & -1 & 1 & 0 & -1\\ \hline
        \textbf{Dynamite} & -1 & 0 & 1 & 1 & 1 & 0 \\ \hline
    \end{tabular}
\end{table}

Without Dynamite, all five moves must be played equally to avoid creating imbalances that the opponent can exploit. Although Dynamite dominates nothing in the payoff matrix in Table \ref{DynLizSpock}, not all moves remain playable after adding Dynamite. The Nash equilibrium excludes both Spock and Lizard. That optimal strategy is shown in Table \ref{rpslsDynStrat}

\begin{table}[!htb]
    \centering
    \caption{Nash Equilibrium for Rock-Paper-Scissors-Lizard-Spock after adding Dynamite.}
    \label{rpslsDynStrat}
    \begin{tabular}{|c|c|c|c|c|c|}
    \hline
    $p_{rock}$ & $p_{paper}$ & $p_{scissors}$ & $p_{Spock}$ & $p_{lizard}$ & $p_{dynamite}$ \\ \hline 
    2/8 & 3/8 & 2/8 & 0 & 0 & 1/8 \\ \hline
    \end{tabular} 
\end{table}

The addition of Dynamite makes two moves unplayable despite not dominating them. Dynamite is also the least used of all the playable moves.

\section{Dynamite vs Lesser Dynamite}
In competition over AI, it is not possible to completely prevent adversaries from accessing AI. Instead, rivals try to develop AI that is more capable than the AI that is available to their adversaries. Weaker adversaries then develop strategies to minimize their disadvantage. In this section, we consider the impact of having more capable Dynamite when the opponent has limited Dynamite rather than when they are completely Dynamite-denied.

To illustrate some dynamics of mutual but unequal access Dynamite, we consider a seven-move one-to-one game.  In this game, one competitor has access to Dynamite that defeats three opponent moves which are arranged as they were for the solved game in section \ref{definingDynamite} and shown in Table \ref{mutual-none}, specifically it defeats moves $\{b_4, b_5, b_6\}$. These are sequentially arranged, meaning that the second two moves defeat the first two moves' countermoves. They also lead up to $b_7$ which will be the move that defeats Dynamite. We will refer to the competitor with this capability as Dynamite-extensive. The Dynamite-limited competitor's Dynamite only defeats two moves. For this example, one of the defeated moves is $a_1$. It is an important move to target because it defeats the $b_7$ move that targets the stronger competitor's Dynamite. We vary the limited-Dynamite's second move. 

\begin{figure}[!htb]
    \centering
    \includegraphics[width=0.5\linewidth]{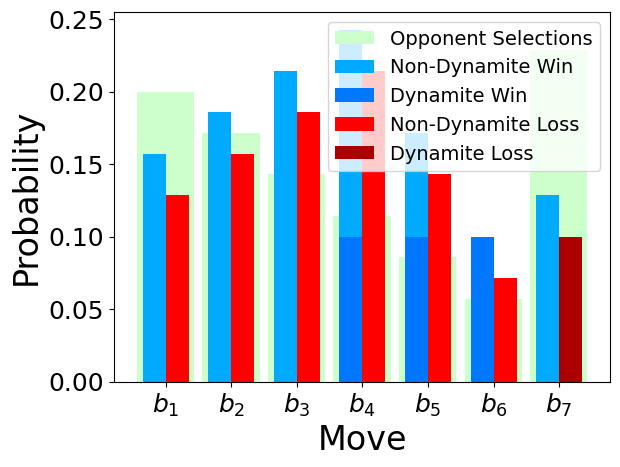}
    \caption{Nash equilibrium strategies for a Dynamite-Denied competitor (green) across all their available moves. For each of those moves, the Dynamite-extensive competitor's equilibrium strategy will result in more wins (blue) than losses (red).}
    \label{fig:mutual-none}
\end{figure}

The payoff matrix in Table \ref{mutual-none} is a baseline for comparison. The Dynamite-limited competitor is fully Dynamite-denied. The equilibrium solution for this game is shown in Figure \ref{fig:mutual-none}. For each Dynamite-denied move, $\{b_1, ..., b_7\}$, the figure shows the probability of wins and losses for the Dynamite-extensive competitor in blue and red, respectively. The difference between the wins and losses is the same in each column, where columns correspond to each possible opponent move. That difference between blue and red is the value of Dynamite, which is 1/35 or about 0.029 in this case. The light green bars show the Nash equilibrium distribution of moves for the Dynamite-denied competitor.

\begin{table}[!htb]
    \centering
    \caption{Payoff matrix for a seven-move one-to-one game between a Dynamite-extensive and a Dynamite-denied competitor.}
    \label{mutual-none} 
    \begin{tabular}{c|c|c|c|c|c|c|c|}
        & $b_1$ & $b_2$ & $b_3$ & $b_4$ & $b_5$ & $b_6$ & $b_7$ \\ \hline
        $a_1$ & -1 & 0 & 0 & 0 & 0 & 0 & 1 \\ \hline 
        $a_2$ & 1 & -1 & 0 & 0 & 0 & 0 & 0 \\ \hline
        $a_3$ & 0 & 1 & -1 & 0 & 0 & 0 & 0 \\ \hline
        $a_4$ & 0 & 0 & 1 & -1 & 0 & 0 & 0 \\ \hline
        $a_5$ & 0 & 0 & 0 & 1 & -1 & 0 & 0 \\ \hline
        $a_6$ & 0 & 0 & 0 & 0 & 1 & -1 & 0 \\ \hline
        $a_d$ & 0 & 0 & 0 & 1 & 1 & 1 & -1 \\ \hline
    \end{tabular}
\end{table}

Table \ref{mutal-b6} is a payoff matrix where the Dynamite-limited competitor can defeat two moves, specifically $a_1$ and $a_6$. The move $a_1$ is the Dynamite-extensive competitor's only option for defeating the Dynamite-limited move $b_7$, which is an important move because it is the one that defeats the Dynamite-extensive competitor's Dynamite.  

\begin{table}[!htb]
    \centering
    \caption{Payoff matrix for a seven-move one-to-one game between a Dynamite-extensive and a Dynamite-limited competitor. The limited Dynamite defeats a move that can be compensated for by the Dynamite-extensive's Dynamite.}
    \label{mutal-b6} 
    \begin{tabular}{c|c|c|c|c|c|c|c|}
        & $b_d$ & $b_2$ & $b_3$ & $b_4$ & $b_5$ & $b_6$ & $b_7$ \\ \hline
        $a_1$ & -1 & 0 & 0 & 0 & 0 & 0 & 1 \\ \hline 
        $a_2$ & 1 & -1 & 0 & 0 & 0 & 0 & 0 \\ \hline
        $a_3$ & 0 & 1 & -1 & 0 & 0 & 0 & 0 \\ \hline
        $a_4$ & 0 & 0 & 1 & -1 & 0 & 0 & 0 \\ \hline
        $a_5$ & 0 & 0 & 0 & 1 & -1 & 0 & 0 \\ \hline
        $a_6$ & -1 & 0 & 0 & 0 & 1 & -1 & 0 \\ \hline
        $a_d$ & 0 & 0 & 0 & 1 & 1 & 1 & -1 \\ \hline
    \end{tabular}
\end{table}

\begin{figure}[!htb]
    \centering
    \includegraphics[width=0.5\linewidth]{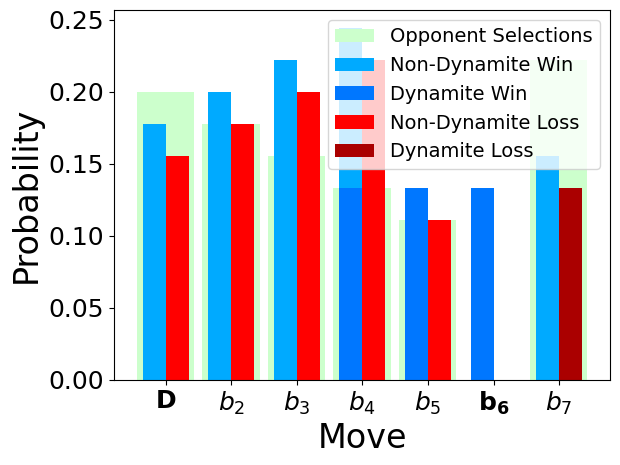}
    \caption{Nash equilibrium strategies for a Dynamite-limited competitor whose Dynamite defeats the first and sixth move competitor (green). For each of those moves, the Dynamite-extensive competitor's equilibrium strategy will result in more wins (blue) than losses (red).}
    \label{fig:unequal-b6}
\end{figure}

Both $a_6$ and $b_5$ are played in the Dynamite-denied case, but in this Dynamite-limited case, $a_6$ falls outside the support. Its vulnerability to the limited Dynamite makes it unplayable given that the move it defeats can also be defeated by Dynamite. As a result of the Dynamite-extensive competitor not playing $a_6$, the Dynamite-limited competitor never plays $b_6$, which is only useful for defeating the no-longer-played $a_6$. That can be seen in Figure \ref{fig:unequal-b6} where there is no green bar, red bar, or non-dynamite bar in the $b_6$ column.

This example shows that the addition of limited Dynamite can make moves obsolete that were not obsolete in the presence of the more capable Dynamite alone. The value of the game decreases from 1/35 to 1/45 or from 0.029 to 0.022. 

The moves $\{a_1, ..., a_7\}$ and $\{b_1, ..., b_7\}$ are not equivalent moves in this one-to-one structure because they do not defeat the same opposing moves and they do not draw against themselves, but in other structures they can be. In that case as well, the addition of limited Dynamite can also decrease the support, making some options obsolete. The moves that become unplayable can be different for the two competitors even if the moves that the limited Dynamite defeats are simply a subset of the moves that the extensive Dynamite defeats. An example illustrating how adding limited Dynamite can make different moves unplayable among competitors that have equivalent sets of non-Dynamite moves is shown in Appendix \ref{appendixMatchedMoves}.

The competition is much different if the limited Dynamite wins against $a_1$ and $a_3$ instead of $a_1$ and $a_6$. That payoff matrix is shown in Table \ref{mutual-b3}. In that case, the defeated move $a_3$ is not one that the extensive Dynamite can compensate for directly. The $a_3$ move is also more threatening to the limited Dynamite, so if the limited Dynamite defeats it rather than $a_6$, then the limited Dynamite is helping to compensate for its own weaknesses.

\begin{figure}[!htb]
    \centering
    \includegraphics[width=0.5\linewidth]{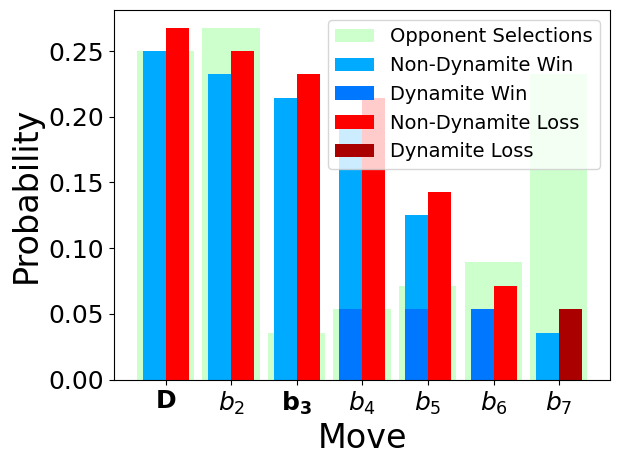}
    \caption{Nash equilibrium strategies for a Dynamite-limited competitor whose Dynamite defeats the first and third move (green). For each of those moves, the Dynamite-extensive competitor's equilibrium strategy will result in more losses (red) than wins (blue).}
    \label{fig:unequal-b3}
\end{figure}

As a result, all moves are within the support, but the Dynamite-limited competitor is in an overall better position than the Dynamite-extensive competitor. The Dynamite-extensive value is \textit{negative} 1/56 or about -0.018. The Dynamite-limited competitor expects to win despite the extensive Dynamite defeating three moves compared to limited Dynamite's two. That can be visualized as in Figure \ref{fig:unequal-b3} where there are red, blue, and green bars for all moves, but where the red bars are taller than the blue ones.

\begin{table}[!htb]
    \centering
    \caption{Payoff matrix for a seven-move one-to-one game between a Dynamite-extensive and a Dynamite-limited competitor. The limited Dynamite defeats a critical move and contributes to its own defense resulting in more expected value.}
    \label{mutual-b3} 
    \begin{tabular}{c|c|c|c|c|c|c|c|}
        & $b_d$ & $b_2$ & $b_3$ & $b_4$ & $b_5$ & $b_6$ & $b_7$ \\ \hline
        $a_1$ & -1 & 0 & 0 & 0 & 0 & 0 & 1 \\ \hline 
        $a_2$ & 1 & -1 & 0 & 0 & 0 & 0 & 0 \\ \hline
        $a_3$ & -1 & 1 & -1 & 0 & 0 & 0 & 0 \\ \hline
        $a_4$ & 0 & 0 & 1 & -1 & 0 & 0 & 0 \\ \hline
        $a_5$ & 0 & 0 & 0 & 1 & -1 & 0 & 0 \\ \hline
        $a_6$ & 0 & 0 & 0 & 0 & 1 & -1 & 0 \\ \hline
        $a_d$ & 0 & 0 & 0 & 1 & 1 & 1 & -1 \\ \hline
    \end{tabular}
\end{table}

We start from a symmetric game where each move is interchangeable with any other. From that starting point, we can provide each competitor with an additional move that defeats different numbers of opposing moves. There are cases where the more limited of the two new additions, with respect to how many opponent capabilities they defeat, can be the more valuable addition.

%\section{Future work}
%Revealing your capabilities. Should you ever play strategies that include moves that would be out of domain to avoid revealing the capabilities of dynamite?

%Progressive improvements. What are the dynamics that occur as dynamite becomes progressively more capable rather than capable all at once. This could be done as here with either wins, losses, or draws. It could also be done by having payoffs be real numbers instead of -1, 0, and 1 then allowing the payoffs for dynamite to increase. 

%Dynamite that loses to many things.

%Dynamite in other configurations. 

%The conditions and prevalence for weakly dominated strategies.

%Increasing capability by selecting the maximally valuable next addition.

\section{Conclusions}
We have considered the model of adding a new more capable Dynamite move to the simple game of Rock-Paper-Scissors to illustrate several aspects of disruption in competition. We draw comparisons to one mechanism by which AI promises to provide advantage--by adding additional high-capability moves or options. 

The Rock-Paper-Scissors-Dynamite model illustrates some ways that a new addition can disrupt a competition without adding value and without changing expected outcomes. It also shows how the disruptive addition might provide less value than anticipated, especially for large games that include many possible moves. And because the value that it does provide does not come from the new move's usage, which is typically low, it suggests the need for alternative value assignments and approaches to pricing as new technology or AI is introduced to some types of competition.

There are also several ways that prior moves can become unplayable, perhaps illuminating mechanisms of technological obsolescence. Moves can be dominated or they can be removed from support. In the Rock-Paper-Scissors model, dominance is rare but a new addition can make many moves strategically unplayable. Still, many moves can remain playable even if the only opposing activity they target is also targeted by the new addition. In fact, those remaining moves can increase in usage, sometimes dramatically. These dynamics illustrate how some technologies, tasks, and professions could become obsolete, and how some others that might seem redundant might increase in importance as AI introduces new highly-capable options.

Providing a new move of lesser capability to a weaker competitor can make some prior moves strategically unplayable even if it only adds capabilities that the stronger competitor already had. Further, in some cases, the more limited addition, that defeats fewer opposing actions while losing to the same number, can be more valuable than the more capable one. 

We found many of these dynamics non-intuitive and initially puzzling. We hope that this model can help others to anticipate some dynamics of competition that would otherwise be surprising. We also hope that by demonstrating how adding new capabilities can provide limited, or even negative value, technology developers and integrators will be able to identify and avoid pitfalls. We hope that they can push products and competitions to be disanalogous from this simple model so as to produce technologies that are valuable  rather than simply ones that are highly-capable.

\section*{Acknowledgments}
The author would like to thank several people for useful discussions, comments, and feedback: Josh Baron, Katherine Quinn, Micah Musser, John Bansemer, Igor Mikolic-Torreira, Jessica Ji, and Joseph Emmens.

\bibliography{references}

\appendix

\section{Nash Equilibrium for Dynamite in Expanded One-to-One Rock-Paper-Scissors}
\label{appendixExpandedOtoO}

At the Nash equilibrium, every move has the same expected value, given the distribution of opponent choices. The payoff matrix then becomes a coefficient matrix for a system of equations. The payoff matrix described in section \ref{definingDynamite}, defines two regions. One region is for moves that dynamite draws against, the other is for those that it defeats. The region where dynamite does not defeat opponent moves applies for $b_1$ to $b_{k-1}$ as in equation \ref{1tok}. For moves $b_k$ to $b_m$, equation \ref{ktom} applies. Here we use the $m$ strategically playable moves rather than the $n$ total available moves. If some moves fall outside the support then $m$ can be less than $n$.

Two regions:
\begin{subequations}
\begin{gather}
    \label{1tok}
    p_{i+1} - p_{i} = p_{i+2}-p_{i+1}, 1 < i < k-1\\
    \label{ktom}
    p_{i+1} - p_{i} + p_d = p_{i+2}-p_{i+1} + p_d, k-1 \le i < m 
\end{gather}
\end{subequations}

The solution in both of these regions has the same functional form but different parameter values as described in equation \ref{funcForm}
\begin{subequations}
\label{funcForm}
\begin{gather}
    p_{i} = C_{1}+iC_{2}, 1 < i < k-1\\
    p_{i} = C_{3}+iC{4}, k-1 \le i < m
\end{gather}
\end{subequations}

These four $C$ parameters can be determined according to four conditions. There are two interface conditions at $k-1$ as shown in equations \ref{continuity} and \ref{derivativeContinuity}. There is the normalization that all probabilities sum to one, shown in equation \ref{normalization}. And there is the cyclic condition from equating the values of the first and last opponent moves shown in equation \ref{cyclical}.

\begin{subequations}
\begin{gather}
    \label{continuity}
    p_{k-1} = p_{k-1} \\
    \label{derivativeContinuity}
    2p_{k-1} = p_{k} + p_{k-2} + p_{d} \\
    \label{normalization}
    \sum_{i=1}^{m-1}p_i + p_d = 1 \\
    \label{cyclical}
    2p_1 = p_2 + p_d, p_d = \frac{p_1 + p_{m-1}}{2}
\end{gather}
\end{subequations}

Those conditions result in the $C$ values as provided in equation \ref{Cvalues}.

\begin{subequations}
\label{Cvalues}
\begin{gather}
    C_1 = \frac{2}{k(m-k+2)} \\
    C_2 = \frac{2(m-k)}{km(m-k+2)} \\
    C_3 = \frac{2}{m-k+2} \\
    C_4 = -\frac{2}{m(m-k+2)}
\end{gather}
\end{subequations}

Those $C$ values determine the probabilities of moves provided in equation \ref{probs}.

\begin{subequations}
\label{probs}
\begin{gather}
    p_i = \frac{2m+2i(m-k)}{km(m-k+2)}, i\le k-1\\
    p_i = \frac{2(m-i)}{m(m-k+2)} , k-1 \le i < m\\
    p_d = \frac{2}{k(m-k+2)} 
\end{gather}
\end{subequations}

The value of the game is $C_2$ as shown again in equation \ref{value}.

\begin{subequations}
\label{value}
\begin{gather}
    v = C_2 =  \frac{2(m-k)}{km(m-k+2)} 
\end{gather}
\end{subequations}

Now we will calculate the number of playable moves to determine whether the total number of available moves $n$ should be used or if $m$ should be capped at a lesser number of moves. 

The Dynamite-holder can defeat some opponent moves using either Dynamite or non-Dynamite moves. They will choose which non-Dynamite moves to retain so as to maximize the value. We will count the moves that Dynamite beats using the index $j$ such that $m = k-1+j$ where $j$ maximizes the value $v$. 

\begin{subequations}
\label{valueDerivative}
\begin{gather}
    \partial v / \partial j = \frac{-2kj^2 + 4kj + 2(2k^2 -k)}{[k(k-1+j)(j+1)]^2} 
\end{gather}
\end{subequations}

Setting the first derivative to zero to find the maximum results in equation \ref{maxJ} for the optimal $j$. Here we leave $j$ as a real number despite being restricted to integers in the game. The particular game solution is of little interest, the simplicity of the square root relationship is valuable for intuition about how many moves become strategically unplayable with the addition of Dynamite.

\begin{subequations}
\label{maxJ}
\begin{gather}
    j = \sqrt{2k}
\end{gather}
\end{subequations}

If there are sufficient available moves that the Dynamite-holder's strategy is maximized by avoiding some options, then the value for $m$ can be replaced in terms of $k$ as shown in equation \ref{valInK}

\begin{subequations}
\label{valInK}
\begin{gather}
    v = \frac{2(\sqrt{k}-1)}{2k^2 + k^{5/2} - k} 
\end{gather}
\end{subequations}

If $n$ is large enough that some moves are outside of support, then as $k$ becomes large, the value of the game scales approximately with the inverse square of $k$ as in equation \ref{valLargeKappendix}.

\begin{subequations}
\label{valLargeKappendix}
\begin{gather}
    v \approx 2k^{-2}
\end{gather}
\end{subequations}

\section{Limited Dynamite in a Matched-Moves One-to-One Competition}
\label{appendixMatchedMoves}

A payoff matrix where each competitor's moves are identical such that each move draws against itself, defeats the same move and loses to the same move is shown in Table \ref{matched-none}.

\begin{table}[!htb]
    \centering
    \caption{Payoff matrix for a seven-move one-to-one game where each competitor's moves are equivalent.}
    \label{matched-none} 
    \begin{tabular}{c|c|c|c|c|c|c|c|}
        & $b_d$ & $b_2$ & $b_3$ & $b_4$ & $b_5$ & $b_6$ & $b_7$ \\ \hline
        $a_1$ & 0 & 1 & 0 & 0 & 0 & 0 & -1 \\ \hline 
        $a_2$ & -1 & 0 & 1 & 0 & 0 & 0 & 0 \\ \hline
        $a_3$ & 0 & -1 & 0 & 1 & 0 & 0 & 0 \\ \hline
        $a_4$ & 0 & 0 & -1 & 0 & 1 & 0 & 0 \\ \hline
        $a_5$ & 0 & 0 & 0 & -1 & 0 & 1 & 0 \\ \hline
        $a_6$ & 0 & 0 & 0 & 0 & -1 & 0 & 1 \\ \hline
        $a_7$ & 1 & 0 & 0 & 0 & 0 & -1 & 0 \\ \hline
    \end{tabular}
\end{table}

We will presume that both players develop Dynamite that defeats the first move, loses to the sixth, and draws against itself. Then one player creates more extensive Dynamite that also defeats the second and third moves. The resulting payoff matrix is shown in Table \ref{matched-baseline}.

\begin{table}[!htb]
    \centering
    \caption{Payoff matrix for a seven-move one-to-one game against very limited Dynamite that only defeats move one.}
    \label{matched-baseline} 
    \begin{tabular}{c|c|c|c|c|c|c|c|}
        & $b_d$ & $b_2$ & $b_3$ & $b_4$ & $b_5$ & $b_6$ & $b_d$ \\ \hline
        $a_1$ & 0 & 1 & 0 & 0 & 0 & 0 & -1 \\ \hline 
        $a_2$ & -1 & 0 & 1 & 0 & 0 & 0 & 0 \\ \hline
        $a_3$ & 0 & -1 & 0 & 1 & 0 & 0 & 0 \\ \hline
        $a_4$ & 0 & 0 & -1 & 0 & 1 & 0 & 0 \\ \hline
        $a_5$ & 0 & 0 & 0 & -1 & 0 & 1 & 0 \\ \hline
        $a_6$ & 0 & 0 & 0 & 0 & -1 & 0 & 1 \\ \hline
        $a_d$ & 1 & 1 & 1 & 0 & 0 & -1 & 0 \\ \hline
    \end{tabular}
\end{table}

The Dynamite-extensive competitor expects to win with a value of 1/28 or 0.0357 and all moves are played for both competitors as shown in Table \ref{matched-baseline-strat}.

\begin{table}[!htb]
    \centering
    \caption{Strategies for both competitors in a competition between extensive dynamite and very limited dynamite that only defeats move one.}
    \label{matched-baseline-strat}
    
    Dynamite-Extensive \\
    \begin{tabular}{|c|c|c|c|c|c|c|}
    \hline
    $p_1$ & $p_2$ & $p_3$ & $p_4$ & $p_5$ & $p_6$ & $p_d$ \\ \hline 
    0.107 & 0.0893 & 0.196 & 0.179 & 0.161 & 0.143 & 0.125 \\ \hline
    \end{tabular} 

    \begin{tabular}{c}
    \end{tabular}

    Dynamite-Limited \\
    \begin{tabular}{|c|c|c|c|c|c|c|}
    \hline
    $p_1$ & $p_2$ & $p_3$ & $p_4$ & $p_5$ & $p_6$ & $p_d$ \\ \hline  
    0.0357 & 0.179 & 0.0714 & 0.214 &  0.107 & 0.25 & 0.143 \\ \hline
    \end{tabular}
\end{table}

If the limited-Dynamite improves to defeat both moves one and two, the payoff matrix is shown in Table \ref{matched-limited}. In this case, the limited Dynamite's capabilities is a subset of the extensive Dynamite.

\begin{table}!htb]
    \centering
    \caption{Payoff matrix for a seven-move one-to-one game where the limited Dynamite's capabilites are a subset of the extensive Dynamite.}
    \label{matched-limited} 
    \begin{tabular}{c|c|c|c|c|c|c|c|}
        & $b_d$ & $b_2$ & $b_3$ & $b_4$ & $b_5$ & $b_6$ & $b_d$ \\ \hline
        $a_1$ & 0 & 1 & 0 & 0 & 0 & 0 & -1 \\ \hline 
        $a_2$ & -1 & 0 & 1 & 0 & 0 & 0 & -1 \\ \hline
        $a_3$ & 0 & -1 & 0 & 1 & 0 & 0 & 0 \\ \hline
        $a_4$ & 0 & 0 & -1 & 0 & 1 & 0 & 0 \\ \hline
        $a_5$ & 0 & 0 & 0 & -1 & 0 & 1 & 0 \\ \hline
        $a_6$ & 0 & 0 & 0 & 0 & -1 & 0 & 1 \\ \hline
        $a_d$ & 1 & 1 & 1 & 0 & 0 & -1 & 0 \\ \hline
    \end{tabular}
\end{table}

The Dynamite-extensive competitor expects to win with a value of 1/33 or 0.03. That is a small reduction in value for the extensive dynamite, moving from one extra victory in 28 games to one extra victory in 33 games. The more qualitative impact is that one move is removed from the support for each competitor and that they are different moves. Move one becomes unplayable for the Dynamite-limited competitor and move two becomes unplayable for the Dynamite-extensive player as shown in Table \ref{matched-limited}

\begin{table}!htb]
    \centering
    \caption{Strategies for both competitors in a competition where the limited Dynamite's capabilities are a subset of the extensive Dynamite.}
    \label{matched-limited}
    
    Dynamite-Extensive \\
    \begin{tabular}{|c|c|c|c|c|c|c|}
    \hline
    $p_1$ & $p_2$ & $p_3$ & $p_4$ & $p_5$ & $p_6$ & $p_d$ \\ \hline 
    0.09 & 0 &  0.24 & 0.15 & 0.21 & 0.12 & 0.18 \\ \hline
    \end{tabular} 

    \begin{tabular}{c}
    \end{tabular}

    Dynamite-Limited \\
    \begin{tabular}{|c|c|c|c|c|c|c|}
    \hline
    $p_1$ & $p_2$ & $p_3$ & $p_4$ & $p_5$ & $p_6$ & $p_d$ \\ \hline  
    0 & 0.18 & 0.09 & 0.21 &
       0.12 & 0.24 & 0.15 \\ \hline
    \end{tabular}
\end{table}

\end{document}